\documentclass[]{llncs}

\usepackage{tikz}
\usepackage{fontawesome}
\usepackage{ amssymb }
\usepackage[]{algorithm2e}
\usetikzlibrary{calc,shapes,arrows,automata,decorations.pathmorphing, shapes.multipart}

\newcommand{\datanode}[3]{
	\node[minimum width=2.5cm,minimum height=1cm, align=center] (#1)  at #2 {#3};
	
	\draw [listingborder]
	(#1.north west)
	-- ([xshift=-\cornerlength]#1.north east)
	-- ([yshift=-\cornerlength]#1.north east)
	-- (#1.south east)
	-- (#1.south west)
	-- cycle;
	\draw [listingborder]
	([yshift=-\cornerlength]#1.north east)
	-| ([xshift=-\cornerlength]#1.north east);
}

\title{Consistency Checking of Functional Requirements}
\author{Simone Vuotto}

\institute{Universit\`a degli Studi di Genova\\\email{simone.vuotto@edu.unige.it}
	\and Universit\`a degli Studi di Sassari\\\email{svuotto@uniss.it}}

\begin{document}

\maketitle

\begin{abstract}
	
Requirements are informal and semi-formal descriptions of the expected behavior of a system. They are usually expressed in the form of natural language sentences and checked for errors manually, \textit{e.g.}, by peer
reviews. Manual checks are error-prone, time-consuming and not scalable.
With the increasing complexity of cyber-physical systems and the need of operating in safety- and security-critical environments, it became essential to automatize the consistency check of requirements and build artifacts to help system engineers in the design process.

\end{abstract}


\section{Introduction}

The assessment of requirements is an important yet costly and complex task, still largely carried manually. The Requirements Engineering (RE)\cite{nuseibeh2000requirements} research field aims at developing tools and techniques to analyze and handle requirements in a more efficient and automatic way. One of the main challenges is to evaluate requirements \textit{consistency}: informally, it means detecting errors, missing information and deficiencies that can compromise the interpretation and implementation of the intended system behavior.  
At a syntactic level, this may involve the check for compliance with standards and guidelines, such as the use of a restricted grammar and vocabulary. We call this task \textit{Compliance Checking}.

However, most of the inconsistencies reside at a semantic level, \textit{i.e.} in their intended meaning. This call for an interpretation and reasoning of requirements semantics. The formalization and translation of requirements into a formal representation is an interesting and open research question.
A recurrent solution in the literature is the use of Property Specification Patterns (PSPs), first introduced by~\cite{dwyer1999}. PSPs provide a direct mapping from English-like structured natural languages to one or more logics. A survey of all available patterns and their translation has been made by~\cite{autili2015aligning}.
Other approaches, like \cite{ghosh2016arsenal}, employ Natural Language Processing
techniques to extract the representation directly from fully natural language requirements. 

Given the set of requirements represented in a formal logic, the main research question is what kind of reasoning we can employ and how to do that. We formally define this task \textit{Consistency Checking} analysis \cite{heitmeyer1996automated}. 
Consistency Checking can range from simple variables type and domain checks to more complex activities, like the evaluation of the intended system behavior over time. In particular, we are interested in checking if the set of requirements together ``make sense'', namely answering the question: 

\begin{center}
\textit{Given the set of requirements, does a system exist that can satisfy them all at the same time?}
\end{center}

The choice of which logic to use is a key research question and it largely affects the reasoning power and the kind of requirements that can be formalized: qualitative, real-time and/or probabilistic. 
We decided to use Linear Temporal Logic (LTL)\cite{pnueli1977temporal} because it is widely used in the literature and it has a good balance between expressiveness and complexity. 
In particular, answering the aforementioned question can be easily translated in a LTL satisfiability check, largely studied and with many efficient tools available~\cite{rozier2010ltl}.

The satisfiability check in turn brings other two research questions: 
\begin{itemize}
	\item \textit{Vacuity Check}: if the formula is satisfiable, is it satisfiable in a meaningful way? For example, the linear temporal logic (LTL) specification $\Box (msg \rightarrow \Diamond rcv)$ (``every message is eventually received'') is satisfied vacuously in a model with no messages, probably not the expected behavior.
	\item  \textit{Inconsistent Requirements Explanation}: if the requirements are inconsistent, which is the minimum set of them that create the inconsistency?
	The number of requirements may be really large, but only few of them making the system unfeasible.	
\end{itemize} 

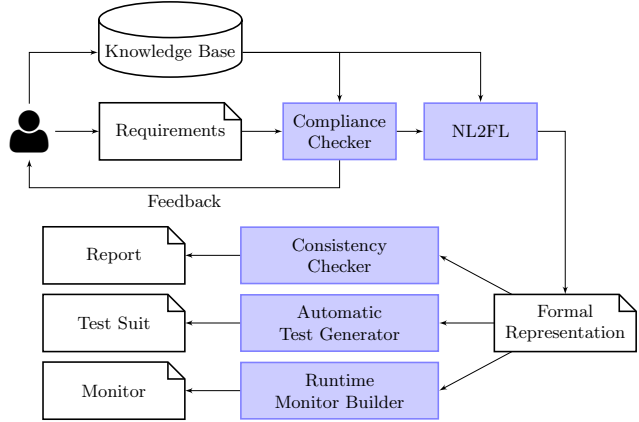
\begin{figure}[t]
	\begin{center}
		\scalebox{0.75}{\begin{tikzpicture}

\newlength{\cornerlength}
\setlength{\cornerlength}{0.3cm}

\tikzset{
	tool/.style = {rectangle, draw=blue!50, fill=blue!20, thick, minimum width=2cm,
		minimum height = 1cm, align=center},
	listingborder/.style={thick},
	db/.style={cylinder, shape border rotate=90, aspect=0.25, thick, draw},
	arrow/.style={draw,-latex'}
}

\node[] (user) at(0, 0) {\Huge\faUser};

\node[db] (kb) at (2.5, 1.4) {Knowledge Base};

\datanode{req}{(2.5, 0)}{Requirements};

\node[tool] (compliance) at (5.5, 0) {Compliance \\ Checker};

\datanode{fl}{(9.5, -3.4)}{Formal\\Representation};

\node[tool] (nl2fl) at (8, 0) {NL2FL};

\node[tool, minimum width=3.5cm] (cc) at (5.5, -2.2) {Consistency \\ Checker};
\node[tool, minimum width=3.5cm] (atg) at (5.5, -3.4) {Automatic\\ Test Generator};
\node[tool, minimum width=3.5cm] (rmb) at (5.5, -4.6) {Runtime\\ Monitor Builder};

\datanode{report}{(1.5, -2.2)}{Report};
\datanode{ts}{(1.5, -3.4)}{Test Suit};
\datanode{monitor}{(1.5, -4.6)}{Monitor};

\path[arrow] (user) -- (req);
\path[arrow] (req) -- (compliance);
\path[arrow] (user) -- (0, 1.4) -- (kb);
\path[arrow] (kb) -- (5.5, 1.4) -- (compliance);
\path[arrow] (kb) -- (8, 1.4) -- (nl2fl);
\path[arrow] (compliance) -- (5.5, -1) -- node[below] {Feedback} (0, -1) -- (0, -0.5);
\path[arrow] (compliance) -- (nl2fl);
\path[arrow] (nl2fl.east) -- (9.5, 0) -- (fl);
\path[arrow] (fl) -- (cc.east);
\path[arrow] (fl) -- (atg.east);
\path[arrow] (fl) -- (rmb.east);
\path[arrow] (cc) -- (report);
\path[arrow] (atg) -- (ts);
\path[arrow] (rmb) -- (monitor);

\end{tikzpicture}}
	\end{center}
	\caption{General framework of the requirement analysis tool}
	\label{fig:main_framework}
\end{figure}

Finally, the formalization of requirements and the consistency checking are enablers for other tasks we would like to tackle in this Ph.D. project, namely the automatic  generation of test suites and runtime monitors. The full overview of the tool that we are designing is depicted in Figure \ref{fig:main_framework}. We are now focusing on the NL2FL
and Consistency Checker modules.

\section{Consistency of Property Specification Patterns}

Our first contribution~\cite{narizzano2018consistency}, developed in the context of the H2020 CERBERO European Project~\cite{cerbero}, presented a tool for the consistency checking of qualitative
requirements expressed in form of PSPs with constrained numerical signals. An example of requirement that we can handle is:
\begin{center}
	\emph{Globally, it is always the case that if $proximity\_sensor < 20$ holds, then $arm\_idle$ eventually holds.} 
\end{center}
We first translate every requirement $r_i \in R$ in LTL$(\mathcal{D}_C)$, an extension of LTL over a constraint system $D_C$ = ($\mathbb{R}$,$<$,$=$), with atomic constraints of the form $x < c$
and $x = c$ (where  $c \in \mathbb{R}$ is a constant real number and `$<$'' and ``$=$'' 
have the usual interpretation).
We then show how the new problem can be reduced to LTL satisfiability.
Let $X(\phi)$ be the set of numerical variables and $C(\phi)$ be the set of
constants that occur in $\phi$. We compute:
\begin{itemize}
	\item the LTL$(\mathcal{D}_C)$ formula $\phi_i$ for every requirement $r_i \in R$;
	\item the conjunctive formula $\phi = \phi_1 \wedge ... \wedge \phi_n$;
	\item a set $M_x(\phi)$ of boolean propositions representing possible values of $x \in X(\phi)$;
	\item the formula $Q_M$ encoding the constraints over $M_x(\phi)$ $\forall x \in X(\phi)$;
	\item the formula $\phi'$ that substitute all $x \in X(\phi)$ in $\phi$ with a set of boolean propositions from $M_x(\phi)$;
\end{itemize}

Given the LTL($\mathcal{D}_C$) formula $\phi$ over the set of Boolean
atoms $Prop$ and the terms $C(\phi) \cup X(\phi)$
we have that $\phi$ is satisfiable if and only if the LTL formula
$\phi_M \rightarrow \phi'$ is satisfiable. This result is important because it shows that LTL($\mathcal{D}_C$) is decidable and that we can exploit state-of-the-art LTL model checkers.

In the second part of the paper we translate our encoding in different formats for of-the-shelf model checkers and we compare their performance. We conclude with the scalability analysis and the methodology application to a robotic arm use-case. 

\section{Future work}

In order to reduce the number of errors in the specification, we have partially implemented an algorithm to check the relationship among requirements.
This is a first step to prevent vacuous results, but more work is needed. 

\paragraph{Connected Requirements Check}
Given a set of requirements $R = \{r_1, ..., r_n\}$, we want to check if one or more of them are completely unrelated from the others, meaning that they describe some behaviors that don't interact with the main bulk of the system. This may happen in an underspecified
requirements set or for some spelling errors (\textit{e.g.} $armidle$ is written in place of $arm\_idle$ in the previous example).
To find these faulty requirements we first build the
undirected graph $G = (V, E)$ representing the connections in $R$, such that:
\begin{itemize}
	\item $v_i \in V$ $\forall r_i \in R$;
	\item $(i, j) \in E$ if $X(r_i) \cap X(r_j) \neq \emptyset$ $\forall r_i, r_j \in R, i \neq j$
\end{itemize}

where $X(r_i)$ is set of variables, boolean or numerical, that appear in $r_i$. We then compute all the connected components in $G$. If the number of components in greater then one, we find the smallest one (\textit{i.e.} the component with the lowest number of vertex) and report it to the user.
\\

Currently we are also focusing our attention on the Inconsistent Requirements Explanation problem. We implemented a simple algorithm that iterate over all $r_i \in R$ and perform the consistency check on the set $R  \setminus r_i$. We keep $r_i$ in $R$ only if the new set is shown consistent, and we discard it otherwise. The algorithm terminates when all the requirements in the original set are checked. This algorithm effectively find a solution, but it is quite inefficient. Therefore, we are seeking for a better algorithm which exploits the structure of the problem.

Finally, for future works we would also like to both extend the natural language interface with less restrictive constraints and adopt a more expressive logic. In particular, we are interested in probabilistic logics such PCTL, but the consistency checking problem is difficult to define in this case and more research is needed.

\paragraph{Acknowledgments} The research of Simone Vuotto is funded by the EU Commission H2020 Programme under grant agreement N.732105 (CERBERO project).

\bibliographystyle{splncs03}
\bibliography{reference.bib}

\end{document}